\documentclass[]{pasj01}
\draft

\Received{}
\Accepted{}
 
\usepackage{natbib}
 
\begin{document} 

\title{ 
\ Light curve analysis of six totally eclipsing \\ W UMa binaries }


\author{Olivera \textsc{Latkovi{\' c}}}
\altaffiltext{}{Astronomical Observatory, Volgina 7, 11060 Belgrade, Serbia}
\email{olivia@aob.rs}

\author{Atila \textsc{{\v C}eki}}
\email{atila@aob.rs}


\KeyWords{binaries (including multiple): close -- binaries: eclipsing -- stars: fundamental parameters -- stars: individual: 
1SWASP J000437.82+033301.2,
1SWASP J050904.45-074144.4,
1SWASP J195900.31-252723.1,
1SWASP J212454.61+203030.8,
2MASS J21031997+0209339,
2MASS J21042404+0731381}

\maketitle

\begin{abstract}
We analyze multicolor light curves of six totally eclipsing, short-period W UMa binaries and derive, for the first time, their orbital and stellar parameters. The mass ratios are established robustly through an automated q-search procedure that performs a heuristic survey of the parameter space. Five stars belong to the W and one to the A subtype. The mass ratios range from 0.23 to 0.51 and the fillouts from 10 to 15\%. We estimate the ages and discuss the evolutionary status of these objects in comparison with a sample of other short-period W UMa binaries from the literature.
\end{abstract}

\section{Introduction}

W UMa stars are low-mass, late-type contact binaries that have evolved into a stable, long-lived common envelope configuration through the shrinking of the orbit due to magnetic breaking and tidal friction \citep{gazeas08, yildiz14}. They are characterized by orbital periods shorter than a day and similar effective temperatures of the components despite the large differences in their masses (low mass ratios). Traditionally, W UMa stars are divided into the A and W subtypes \citep{binnen70}: the larger, more massive component is also the hotter one in the A-type, and cooler in the W-type stars. A tertiary component may play an important role in the removal of the angular momentum from these systems; indeed, many of W UMa binaries are confirmed members of hierarchical multiple systems \citep{pribulla06, tokovinin06}.

Although these objects have been studied extensively, open questions still exist regarding their evolutionary status and the mechanisms of the mass and energy exchange through the common envelope \citep{webbink03}. W UMa binaries with very short periods near or under the period cutoff at 0.22 days are especially interesting because the cutoff has not yet been fully explained \citep{li19}.

\citet{koen16} published the multicolor light curves of 29 short-period W UMa binaries together with their color indices and spectral types. From this data, we selected six systems that display total eclipses for further study. Their basic properties are listed in Table \ref{tabStars}. Four of them have periods near the 0.22 days period cutoff, and one under it. To our knowledge, none of these stars have been modeled in detail prior to this work. 

\begin{table}
  \tbl{The main characteristics of the selected W UMa stars.}{%
  \begin{tabular}{llccccc}
      \hline
      Nickname & ID & 
      $\alpha_{2000}$\footnotemark[$*$] & 
      $\delta_{2000}$\footnotemark[$*$] & 
      $m_V$\footnotemark[$*$] & 
      Period [d]\footnotemark[$**$] \\
      \hline
      J000437 & 1SWASP J000437.82+033301.2 & 00 04 37.84 & +03 33 01.14 & 14.53 & 0.26150 \\
      J050904 & 1SWASP J050904.45-074144.4 & 05 09 04.45 & -07 41 44.25 & 13.30 & 0.22958 \\
      J195900 & 1SWASP J195900.31-252723.1 & 19 59 00.31 & -25 27 23.1  & 14.64 & 0.23814 \\
      J210319 & 2MASS J21031997+0209339    & 21 03 19.98 & +02 09 33.93 & 15.31 & 0.22859 \\
      J210424 & 2MASS J21042404+0731381    & 21 04 24.04 & +07 31 38.08 & -     & 0.20909 \\
      J212454 & 1SWASP J212454.61+203030.8 & 21 24 54.61 & +20 30 30.8  & 14.35 & 0.22783 \\
      \hline
    \end{tabular}}\label{tabStars}
\begin{tabnote}
\footnotemark[$*$] Taken from the Simbad database ({\it http://simbad.u-strasbg.fr/simbad/}). \\
\footnotemark[$**$] Taken from \citet{norton11} for all stars except J00437 and J195900, for which the periods were improved by \citet{lohr12}. \\
\end{tabnote}
\end{table}

In what follows, we describe the steps taken to prepare the data for light curve analysis (Section \ref{secData}), the q-search procedure (Section \ref{secQS}) and the fine-tuning of the models for each star in our sample (Section \ref{secModels}). The estimates of their absolute parameters and ages are given in Section \ref{secResults} and compared with a sample of other recently studied short-period W UMa binaries.

\section{Data preparation}
\label{secData}

After folding the light curves in orbital phases using the periods from Table \ref{tabStars}, we timed the eclipses by fitting the region around the eclipse with a low-order polynomial. These eclipse timings are given in Table \ref{tabET}. Phased light curves were then normalized by subtracting the magnitude at light maximum from all the points, so that the light curve maximum is always around zero.

\begin{table}
  \tbl{The eclipse timings of the selected W UMa stars.}{%
  \begin{tabular}{llcc}
      \hline
      Nickname & $\rm T_I$ [HJD] & $\rm T_{II}$ [HJD] \\ 
      \hline
      J000437  & 2456881.5209    & 2456881.6508       \\
      J050904  & 2457016.7919    & 2457016.6780       \\
      J195900  & 2456881.5819    & 2456881.7013       \\
      J210319  & 2457277.6999    & 2457277.5830       \\
      J210424  & 2457270.7246    & 2457270.8304       \\
      J212454  & 2457271.8479    & 2457271.7335       \\
      \hline
    \end{tabular}}\label{tabET}
\end{table}

\section{Light curve analysis}
\label{secAnalysis}

The light curve analysis was done using a modernized version of the program by \citet{djur92} generalized for the case of contact configurations \citep{djur98}. The program implements a binary star model based on Roche geometry. An example of its application to contact binaries, with a detailed description of all the model parameters, can be found in \citet{caliskan14}. In the present work, we use the $\alpha$-constrained Nelder-Mead Simplex \citep{takahama03} to fit the synthetic light curves to the observations.

\subsection{The q-search}
\label{secQS}

Since no radial velocity time series have been published for our six stars so far, we conduct a fully automated heuristic q-search to determine the optimal mass ratios. This involves generating 100 models with randomized initial parameters for 100 values of the mass ratio in the range from 0 to 1 (for a total of 10000 random trial models), fitting them to the observations and selecting the mass ratio of the best-fitting one.

For every star, this is done twice: for the A-type configuration (the deeper minimum corresponds to the eclipse of the more massive star) and for the W-type configuration (the deeper minimum corresponds to the eclipse of the less massive star). Here and throughout this work, the more (less) massive star is considered to be the primary (secondary) and is indexed with the number 1 (2). In the case of the W-type configuration, this convention requires phase-shifting the light curve (via the ``phase shift'' model parameter) so that the deeper minimum has the phase 0.5 (whereas with the A-type, it's at phase 0.0, as is usual).

The randomized parameters are: 1) The orbital inclination ($i$), which we sample from a uniform distribution in the range from $45^{\circ}$ to $90^{\circ}$. 2) The filling factor of the primary star, $F_1$, which is a measure of its size relative to the size of its inner critical Roche lobe. In contact binaries, the components share a common envelope described by a single equipotential surface, so the size of the secondary is constrained by the size of the primary via the mass ratio as $R_2/R_1\approx (M_2/M_1)^{0.46}$ \citep{kuiper41}. The values of $F_1$ are sampled from a uniform distribution in the range from 1.001 to a maximum value corresponding to the outer critical Roche lobe. 3) The temperature of the secondary ($T_2$) or the primary ($T_1$) star in the A-type or W-type scenario, respectively. The values are drawn from a uniform distribution between 3000 and 8000 K. 4) Third light, defined as the contribution of constant, uneclipsed flux to the total flux at maximum light: $\ell_3=F_3/(F_{1, max} + F_{2, max} + F_3)$, is added to all models as the preliminary fittings proved it necessary in most cases. The values are taken from a uniform distribution ranging from 0 to 1. Candidate parameters are drawn from such wide ranges because this stage of the analysis was fully automated and applied to all the stars observed by \citet{koen16} prior to the careful inspection of individual light curves.

During the q-search and in subsequent tweaking of the models, the mass ratio is kept fixed to its grid values, while the parameters listed above are adjusted. The albedos ($A_1$ and $A_2$) and gravity darkening exponents ($\beta_1$ and $\beta_2$) are kept fixed to the theoretical values appropriate for each component according to its temperature \citep{vonz,lucy,rucinski69}. The nonsynchronous rotation coefficient, $f=\omega_{rot}/\omega_{orb}$, is constant and equal to 1, as we assume synchronized rotation. The phase shift (with initial values of 0 and 0.5 for A- and W-type configurations, respectively) and the magnitude shift are always adjusted for small translations along the axes.

\begin{figure}
 \begin{center}
  \includegraphics[width=0.9\textwidth]{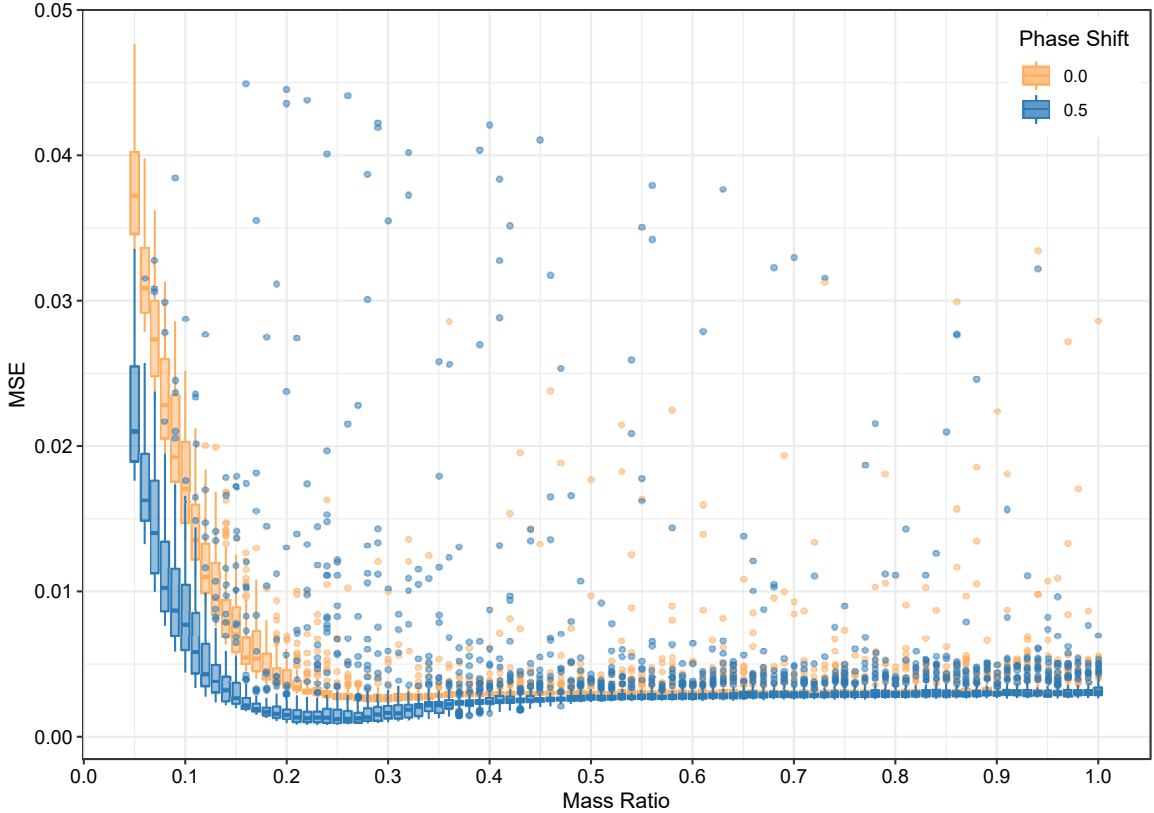} 
 \end{center}
\caption{The details of the q-search for J000437. The box-plots summarize the fit quality of the 100 randomized models at each trial value of the mass ratio. The orange symbols represent the A-type, and the blue, the W-type configurations.}\label{figQSDemo}
\end{figure}

The q-search procedure is illustrated in Fig. \ref{figQSDemo}. As a measure of goodness-of-fit we use the mean squared error, $MSE=\frac{1}{N_{obs}}\sum(O-C)^2$. The values of MSE after fitting the initial, randomized models to the observations are grouped by mass ratios and represented with box-plots. The height of the box corresponds to the interquartile range ($IQR$), and the whiskers to the min/max range of the achieved MSE values. The horizontal line inside the box is the median of the distribution. Outliers, defined as points that lie outside $1.5 \times IQR$, are plotted individually. Each box summarizes the 100 final models for the corresponding mass ratio. 

With this procedure, we not only estimate the most likely value of the mass ratio, but also perform a heuristic search of the parameter space. Thanks to the random sampling of other model parameters, the region near the minimum of a q-search curve indicates the location of the globally best-fitting model, providing a fair certainty that our solutions are unique.

\begin{figure}
 \begin{center}
  \includegraphics[width=0.9\textwidth]{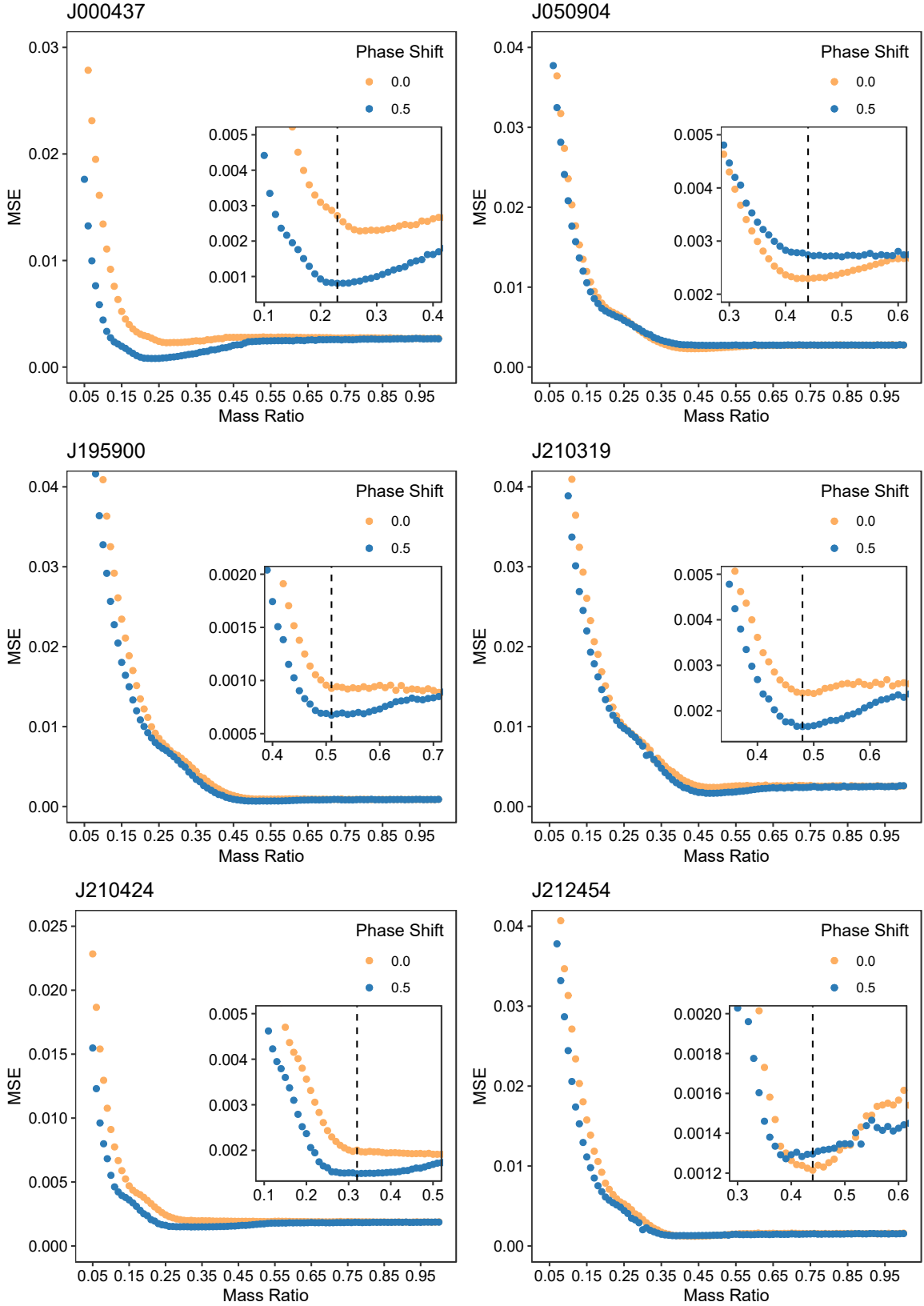} 
 \end{center}
\caption{The q-search curves for all six stars in our sample. The inset zooms in on the region of the q-search curve minimum, and the final mass ratios are marked with vertical dashed lines. As in Fig. \ref{figQSDemo}, the orange symbols show the A-type, and the blue, the W-type configurations.}\label{figQSearch}
\end{figure}

The q-search curves for all the stars in our sample are shown in Fig. \ref{figQSearch}, where we only plot the points with the lowest MSE for each mass ratio for clarity. The insets zoom in on the region of the minimum, and the final mass ratio is indicated with a vertical dashed line.

In all the plots in Figs. \ref{figQSDemo} and \ref{figQSearch}, the orange symbols stand for the A-type (no phase-shift) and the blue ones for the W-type configuration (phase shift of 0.5). Thanks to the total eclipses, the type of the binary is unambiguously determined by the q-search, as for the A-type stars the total eclipse must be in the shallower, and in the W-type systems, in the deeper minimum.

\clearpage

We have performed the automated q-search described above for all the contact binaries observed by \citet{koen16}. Among them, only the six analyzed in this work have q-search curves with definite dips, where a minimum can be established without doubt. The rest, which either have incomplete light curves or display only partial eclipses, have q-search curves that slope monotonically toward unity mass ratio. This is because the orbital inclination cannot be constrained by the shape of the light curve alone unless the eclipses are total. For these objects to be modeled reliably, it would be necessary to also study their radial velocity curves.

For our six totally eclipsing stars, the best-fitting q-search model is selected for further, manual tweaking. These initial models already fit the observations quite well, but not perfectly. The most notable improvements can be made by adding spots to the models, to account for light curve asymmetries. For dark spots, the initial temperature contrast, $T_{spot}/T_{star}$, is 0.8, and for bright spots, 1.2. Whenever possible, we only adjust the longitude of the spot and leave the latitude fixed to zero (equatorial spots). 

\begin{figure}
 \begin{center}
  \includegraphics[width=0.9\textwidth]{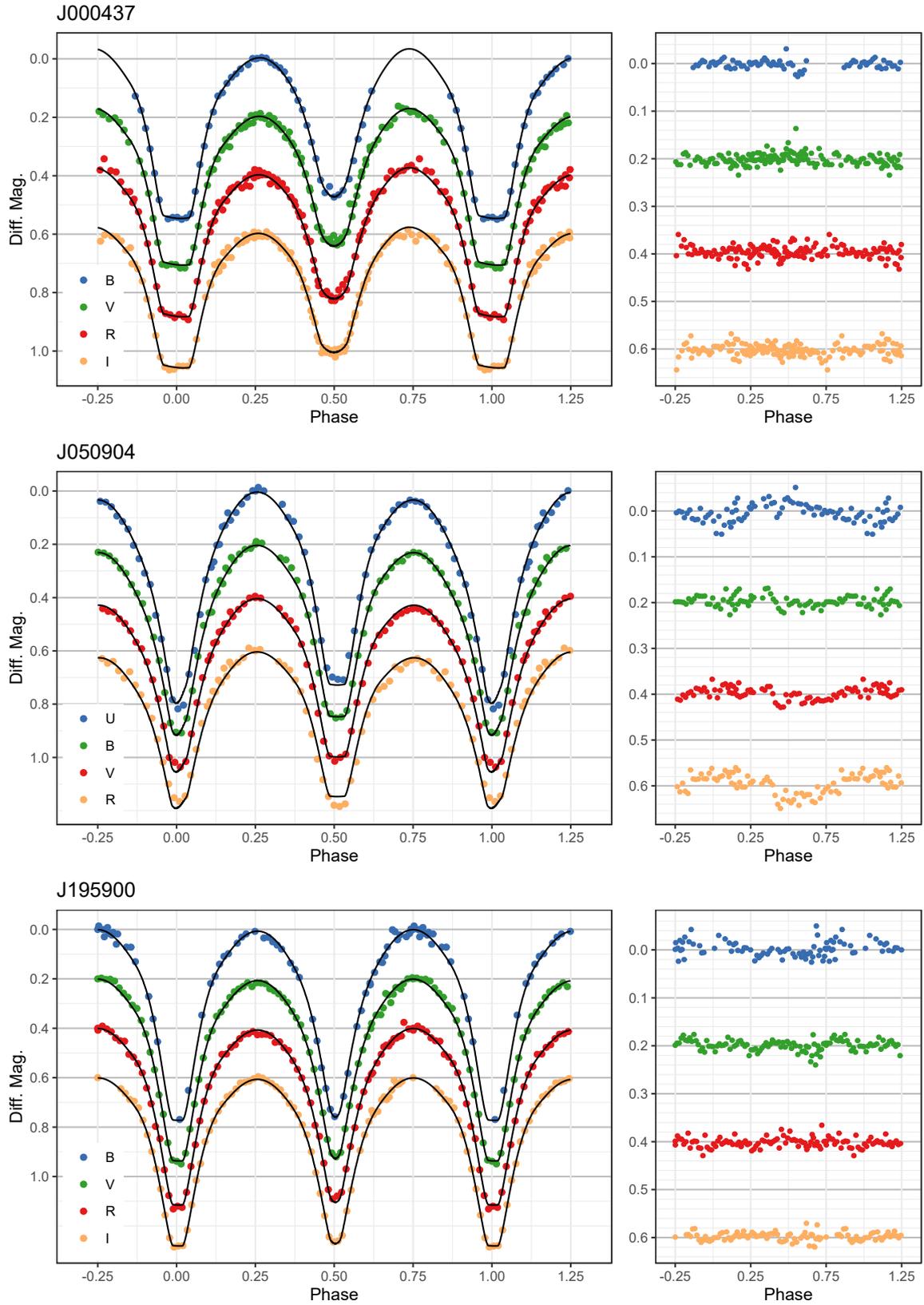} 
 \end{center}
\caption{Left: the observed (points) and synthetic (solid lines) light curves of J000437, J050904 and J195900. Right: the residuals. The \textit{VRI} (in the case of J050904, the \textit{BVR}) filters are offset for clarity by 0.2, 0.4 and 0.6 mag, respectively.}\label{figModels1}
\end{figure}

\begin{figure}
 \begin{center}
  \includegraphics[width=0.9\textwidth]{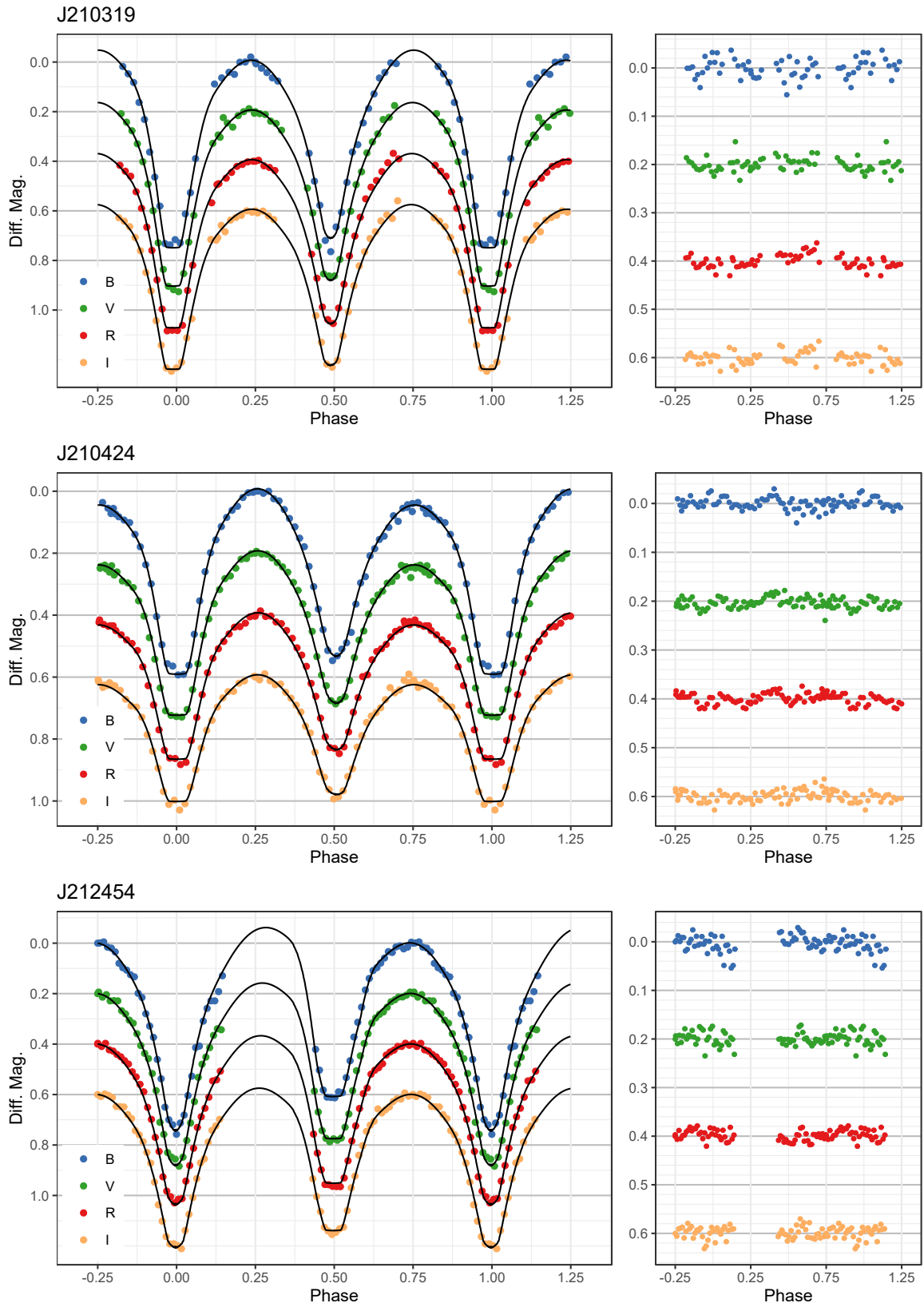} 
 \end{center}
\caption{Same as in Fig. \ref{figModels1}, for J210319, J210424 and J212454.}\label{figModels2}
\end{figure}

\begin{figure}
 \begin{center}
  \includegraphics[width=0.9\textwidth]{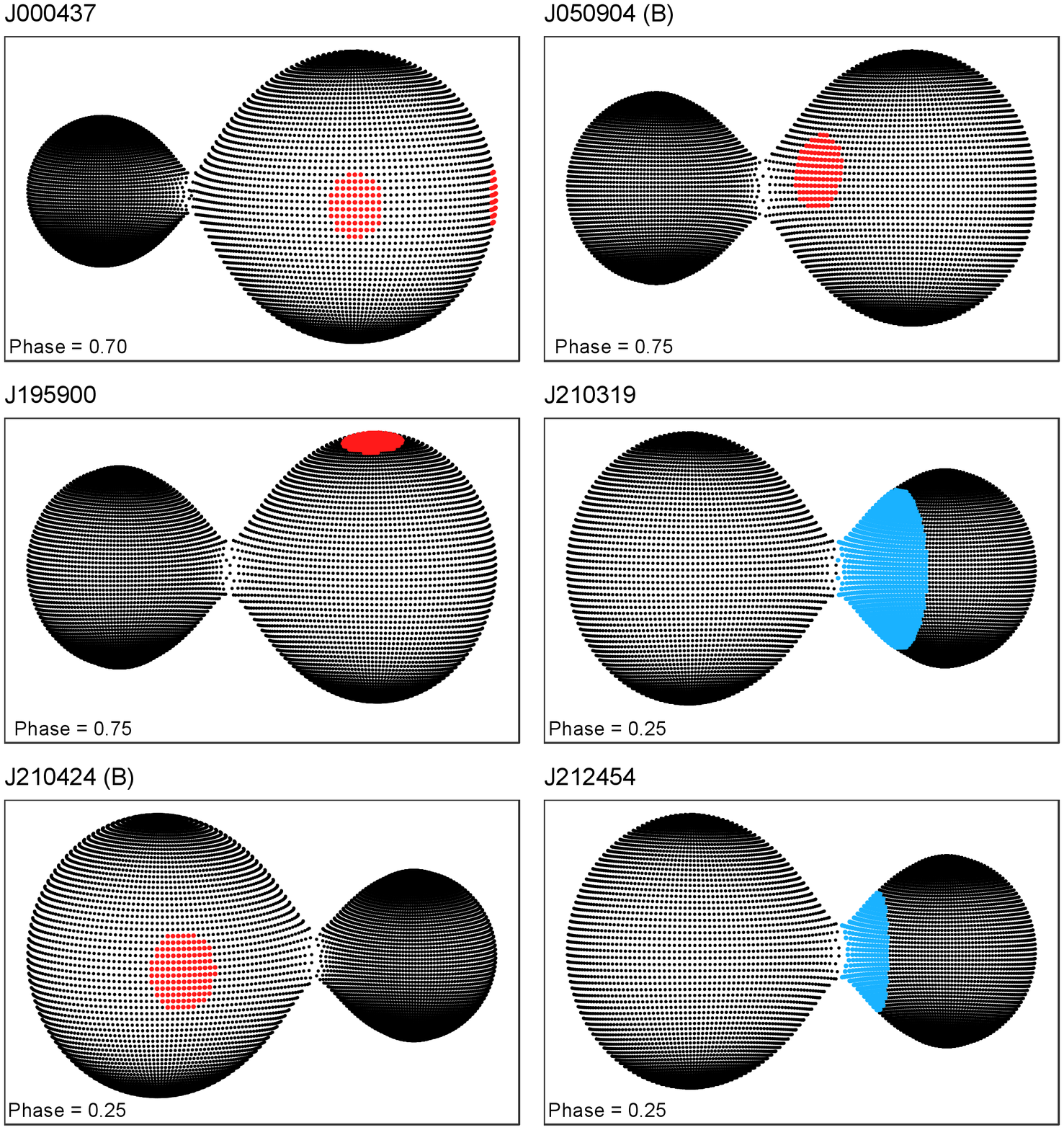} 
 \end{center}
\caption{A visualization of the final models described in Section \ref{secModels}. Dark and bright spots are indicated by red and blue symbols, respectively.}\label{fig3D}
\end{figure}

The improved, final models, are shown in Figs. \ref{figModels1}, \ref{figModels2} and \ref{fig3D} and their parameters are summarized in Table \ref{tabModels1}. In the following paragraphs, we proceed to describe the modeling of each star individually.

\subsection{Final models}
\label{secModels}

\textit{J000437}

\noindent The light curves of J000437 show a total eclipse in the deeper minimum, which means it is a W-type system. The q-search curve confirms this, with the phase-shifted, W-type models achieving systematically better fits to the observations than the A-type models at the same mass ratios. The mass ratio resulting in the best fit is $q=0.23$. It is the lowest mass ratio in the sample. The q-search solution has a small third light contribution of about 3\% in all filters, but it was possible to fit the observations with a model that doesn't require it. The light maximum at phase 0.25 is slightly lower than the one in phase 0.75, indicating surface brightness inhomogeneities that we could model by adding two dark, equatorial spots on the primary star. 
\newline

\noindent \textit{J050904}

\noindent The q-search identifies J050904 as an A-type W UMa binary, but in the final model, the larger, more massive component has the lower temperature, which is the defining property of W-type stars. On closer inspection of the light curves, it can be seen that the total eclipse is shallower in the \textit{U} and \textit{B} filters, about the same depth as the other one in the \textit{V} filter, but \textit{deeper} in the \textit{R} filter. This renders the subtype assignment ambiguous. We adopt the W-type configuration, as that is how we would classify this star according to the final model parameters. 

The mass ratio of J050904 obtained from the q-search is $q=0.44$. Here, the contribution of the third light is significant and wavelength-dependent: from under 1\% in the \textit{U} filter to above 10\% in the \textit{R} filter. We could not find a model that can fit the observed light curve amplitude without the third light, so it is included in the final model too. Also included is a medium-sized dark spot at a low latitude on the primary component, to account for the small asymmetry between the maxima at phases 0.25 and 0.75.

The curious inversion of relative eclipse depths cannot be readily explained by our model. It is possible to reproduce it through the variation of component radii, temperatures or spot parameters with wavelength if, instead of fitting all the filters simultaneously, a separate model is fitted to each filter individually. While this improves the fit to the observations inside the total eclipse, it doesn't lead to results fundamentally different from those shown in Table \ref{tabModels1}. We also tried to model J050904 as near-contact binary, but such models can't achieve fits as good as the contact ones, and don't reproduce the inversion either.
\newline

\noindent \textit{J195900}

\noindent This is a W-type system with a short total eclipse in the deeper minimum. The automated q-search results in the mass ratio of $q=0.51$ and gives a model that is already satisfactory. We could only improve it slightly by adding a dark spot at a relatively high latitude to account for small light curve asymmetries. The third light contribution in this system is small (around 1\%), but significant, as a good fit cannot be achieved when it is eliminated.
\newline

\clearpage

\noindent \textit{J210319}

\noindent This star has partially incomplete light curves, but the gaps in the data are small, so it is possible to fully model the system despite them. The minima are of nearly equal depth, but the q-search indicates that this is another W-type binary, with the total eclipse in the deeper minimum. The mass ratio resulting from q-search is $q=0.48$. The third light contribution is wavelength-dependent, ranging from 3\% in the \textit{B} to 6\% in the \textit{I} filter, and necessary for matching the light curve amplitude. A rather large, bright, equatorial spot of very low temperature contrast located in the ``neck" region near the Largangian point $L_1$ is found to improve the model by accounting for slight light curve asymmetry. Such spots are usually interpreted as anomalies in surface temperature distribution arising from mass and energy transfer \citep[see e.g.][]{lu91, luo15, djur16}\footnote{See also \citet{zhou90}, \citet{oka02} and \citet{stepien09} for an overview of large-scale surface motions in contact binaries and their effects on the energy exchange through the common envelope.}.
\newline

\noindent \textit{J210424}

\noindent The light curves of J210424 have a total eclipse in the deeper minimum, meaning it belongs to the W-type. The mass ratio resulting from the q-search is $q=0.32$, the second lowest in the sample. At the same time, the third light contribution is the highest in the sample: from 10\% in the \textit{B} to 23\% in the \textit{R} filter. A dark, equatorial spot is added to the final model to account for light curve asymmetry. 
\newline

\noindent \textit{J212454}

\noindent The light curves of J212454 are also partly incomplete, missing observations between phases 0.15 and 0.45. However, both minima are fully covered and the shallower one displays a totality. The q-search results in an A-type model with the mass ratio of $q=0.44$ and wavelength-dependent third light contribution between 5 and 8\%. The missing maximum at phase 0.25 rules out the possibility to model asymmetry-inducing spots on the sides of the stars; however, the model could be significantly improved by adding a bright spot to the secondary star in the ``neck" region, similar to the one in the model of J210319.

\clearpage

\begin{table}
\tbl{The parameters of the best-fitting models for the selected W UMa stars.}{
\begin{tabular}{lrrrrrr}
\hline                                                                                          
Star/Quantity         & J000437   & J050904   & J195900   & J210319   & J210424   & J212454   \\
\hline                                                                                          
\hline                                                                                          
$q$                   & 0.23      & 0.44      & 0.51      & 0.48      & 0.32      & 0.44      \\
$i [^{\circ}]$        & 86.67     & 89.51     & 86.13     & 87.89     & 83.25     & 89.01     \\
$f_{over} [\%]$       & 11.64     & 14.30     & 12.92     & 13.47     & 15.77     & 13.42     \\
$T_1 [K]$             & 5068      & 4840      & 5027      & 3927      & 4220      & 4840      \\
$T_2 [K]$             & 5340      & 4933      & 5170      & 4050      & 4450      & 4810      \\
$r_1 [a_{orb}]$       & 0.5167    & 0.4631    & 0.4495    & 0.4550    & 0.4910    & 0.4625    \\
$r_2 [a_{orb}]$       & 0.2669    & 0.3206    & 0.3322    & 0.3274    & 0.2952    & 0.3200    \\
$F_{1,2}$             & 1.0083    & 1.0160    & 1.0158    & 1.0165    & 1.0138    & 1.0148    \\
$\Omega_{1,2}$        & 2.2888    & 2.7194    & 2.7206    & 2.7984    & 2.4787    & 2.7229    \\
$\Omega_{in}$         & 2.3058    & 2.7586    & 2.7586    & 2.8372    & 2.5100    & 2.7586    \\
$\Omega_{out}$        & 2.1601    & 2.4925    & 2.5911    & 2.5494    & 2.3114    & 2.4925    \\
$\ell_3 (U)$          & -         & 0.0077    & -         & -         & -         & -         \\
$\ell_3 (B)$          & -         & 0.0642    & 0.0091    & 0.0296    & 0.0959    & 0.0545    \\
$\ell_3 (V)$          & -         & 0.0915    & 0.0081    & 0.0356    & 0.1245    & 0.0679    \\
$\ell_3 (R)$          & -         & 0.1238    & 0.0065    & 0.0485    & 0.1725    & 0.0802    \\
$\ell_3 (I)$          & -         & -         & 0.0142    & 0.0565    & 0.2312    & 0.0801    \\
\hline                                                                                          
Spot\ 1               & Primary   & Primary   & Primary   & Secondary & Primary   & Secondary \\
\hline                                                                                          
$T_{spot}/T_{star}$   & 0.8205    & 0.8000    & 0.8059    & 1.0323    & 0.8000    & 1.2036    \\
$\sigma$              & 10.92     & 14.52     & 16.97     & 53.11     & 15.36     & 33.46     \\
$\lambda$             & 168.53    & 324.59    & 273.70    & 151.39    & 77.77     & 169.31    \\
$\varphi$             & 0.00      & -0.44     & 77.61     & 0.00      & 0.00      & 0.00      \\
\hline                                                                                          
Spot\ 2               & Primary   &           &           &           &           &           \\
\hline                                                                                          
$T_{spot}/T_{star}$   & 0.80      &           &           &           &           &           \\
$\sigma$              & 11.80     &           &           &           &           &           \\
$\lambda$             & 251.62    &           &           &           &           &           \\
$\varphi$             & 0.00      &           &           &           &           &           \\
\hline                                                                                          
$MSE$                 & 0.0006    & 0.0013    & 0.0006    & 0.0010    & 0.0005    & 0.0007    \\
\hline                                                                                          
\end{tabular}}\label{tabModels1}
\end{table}

\section{Absolute parameters and age}
\label{secResults}

We estimate the absolute parameters (masses, radii and luminosities) of the components of our six stars assuming that the primary is on the main sequence and that the calibrations for normal stars are applicable \citep[see e.g.][]{yakut05}. The primary mass can be estimated from its temperature via the spectral type--temperature and spectral type--mass calibrations; the secondary mass then follows from the mass ratio. Knowing the total mass, we can obtain the orbital separation from the third Kepler law, and calculate the mean radii in solar units. Finally, the luminosities follow from the absolute radii and effective temperatures. In all calculations, we use the calibrations and solar quantities from \citet{allen}.

The results are given in Table \ref{tabAbs}. It can be seen from this listing that the primaries of all the systems have radii and temperatures typical for the main sequence, while the secondaries are oversized and overluminous compared to main sequence stars of the corresponding mass. This is an expected property of W UMa binaries. The differences between component temperatures range between 30 K (J212454) and 270 K (J000437) with the mean at about 150 K, indicating a good thermal contact despite relatively small fillouts (from 10 to 15\%; see Table \ref{tabModels1}).

\begin{table}
\tbl{The absolute parameters and ages of the selected W UMa stars.}{
\begin{tabular}{lrrrrrr}
\hline
Quantity & J000437 & J050904 & J195900 & J210319 & J210424 & J212454 \\ 
\hline
\hline
  $a [R_{\odot}]  $ & 1.72 & 1.62 & 1.73 & 1.43 &  1.37 &  1.62 \\ 
  $M_1 [M_{\odot}]$ & 0.82 & 0.76 & 0.81 & 0.51 &  0.59 &  0.76 \\ 
  $M_2 [M_{\odot}]$ & 0.19 & 0.33 & 0.41 & 0.24 &  0.19 &  0.33 \\ 
  $R_1 [R_{\odot}]$ & 0.89 & 0.75 & 0.78 & 0.65 &  0.67 &  0.75 \\ 
  $R_2 [R_{\odot}]$ & 0.46 & 0.52 & 0.57 & 0.47 &  0.40 &  0.52 \\ 
  $T_1 [K]$         & 5068 & 4840 & 5027 & 3927 &  4220 &  4840 \\ 
  $T_2 [K]$         & 5340 & 4933 & 5170 & 4050 &  4450 &  4810 \\ 
  $L_1 [L_{\odot}]$ & 0.47 & 0.28 & 0.35 & 0.09 &  0.13 &  0.28 \\ 
  $L_2 [L_{\odot}]$ & 0.15 & 0.14 & 0.21 & 0.05 &  0.06 &  0.13 \\ 
  $\log(g)_1$       & 4.45 & 4.57 & 4.57 & 4.52 &  4.56 &  4.57 \\ 
  $\log(g)_2$       & 4.39 & 4.53 & 4.54 & 4.49 &  4.51 &  4.54 \\ 
  $M^{bol}_1$       & 5.61 & 6.18 & 5.94 & 7.40 &  7.01 &  6.18 \\ 
  $M^{bol}_2$       & 6.82 & 6.90 & 6.47 & 7.98 &  7.89 &  7.01 \\ 
  $\tau [\rm Gyr]$  & 9.59 & 9.04 & 6.97 & 9.52 &     - &     - \\ 
\hline
\end{tabular}}\label{tabAbs}
\end{table}

Table \ref{tabAbs} also shows the estimates of age that we have made using the method presented by \citet{yildiz13} and \citet{yildiz14}. The ages seem plausible for J000437 (9.59 Gyr), J050904 (9.04 Gyr), J195900 (6.97 Gyr) and J210319 (9.52 Gyr), but for J210424 and J212454 we could not get estimates shorter than the current age of the universe. As these two stars have the shortest periods in the sample (J212454 is almost exactly at, and J210424 is below the short period cutoff), we suspect that the age estimation method may not be suitable for ultra-short-period objects. The age estimates for the other four stars should also be taken as tentative.

\begin{figure}
 \begin{center}
  \includegraphics{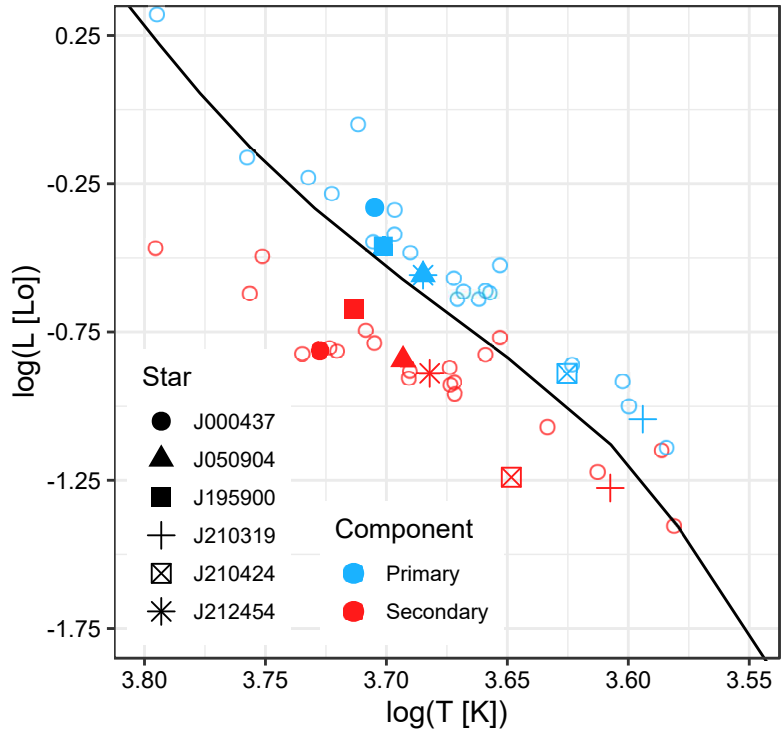} 
 \end{center}
\caption{The selected W UMa binaries on the HR diagram. The primary stars are colored blue, and the secondary, red. Objects from Table \ref{tabCompare} are also shown as smaller, empty circles. The solid line indicates the main sequence extracted from the MIST model archive \citep{MIST1, MIST2}.}\label{figHRD}
\end{figure}

A sample of twenty short-period W UMa binaries from recent studies is shown in Table \ref{tabCompare}. Like our targets, most of these stars have mass ratios under 0.5, primaries of sub-solar masses and radii, and temperature differences under 200 K. Temperatures range quite widely, from 3800 to 6250 K, which corresponds to spectral types between M0 and F8. The spectral types of our stars, estimated by \citet{koen16}, fall in a narrower range: from K5 (J210319) to G5 (J000437). Our targets are placed on a HR diagram together with this sample in Fig. \ref{figHRD}, where we see that they occupy the same region.

\begin{table}
\tbl{A sample of short-period totally eclipsing W UMa binaries from the literature.}{%
\begin{tabular}{lrrrrrrrrr}
\hline
Name & P [d] & q & $M_1 [M_{\odot}]$ & $M_2 [M_{\odot}]$ & $R_1 [R_{\odot}]$ & $R_2 [R_{\odot}]$ & $T_1 [K]$ & $T_2 [K]$ & Ref.\footnotemark[$*$] \\ 
\hline
\hline
Cl* NGC 6121 SAW V66       & 0.2700 & 0.12 & 1.66 & 1.24 & 1.24 & 0.50 & 6238 & 6244 &  (1) \\ 
GSC 4946-0765              & 0.2697 & 0.43 & 0.93 & 0.40 & 0.88 & 0.59 & 5400 & 5639 &  (2) \\ 
EK Com                     & 0.2667 & 0.28 & 1.03 & 0.28 & 0.98 & 0.57 & 5150 & 5290 &  (3) \\ 
CRTS J004259.3+410629      & 0.2664 & 0.28 & 1.19 & 0.33 & 1.02 & 0.58 & 5724 & 5706 &  (4) \\ 
V1197 Her                  & 0.2627 & 0.38 & 0.77 & 0.30 & 0.83 & 0.54 & 4973 & 5113 &  (5) \\ 
PZ UMa                     & 0.2627 & 0.18 & 0.77 & 0.14 & 0.92 & 0.43 & 4972 & 5430 &  (6) \\ 
1SWASP J150822.80-054236.9 & 0.2601 & 0.51 & 1.07 & 0.55 & 0.90 & 0.68 & 4500 & 4500 &  (7) \\ 
CSS J004004.7+385531       & 0.2512 & 0.55 & 0.79 & 0.43 & 0.80 & 0.61 & 4560 & 4560 &  (5) \\ 
1SWASP J064501.21+342154.9 & 0.2486 & 0.48 & 0.70 & 0.30 & 0.76 & 0.55 & 4590 & 4720 &  (9) \\ 
YZ Phe                     & 0.2347 & 0.38 & 0.74 & 0.28 & 0.75 & 0.48 & 4658 & 4908 & (10) \\ 
V1009 Per                  & 0.2341 & 0.36 & 0.87 & 0.31 & 0.86 & 0.47 & 5280 & 5253 & (11) \\ 
1SWASP J044132.96+440613.7 & 0.2282 & 0.64 & 0.70 & 0.44 & 0.72 & 0.59 & 4003 & 3858 & (12) \\ 
1SWASP J093010.78+533859.5 & 0.2277 & 0.40 & 0.86 & 0.34 & 0.79 & 0.52 & 4700 & 4700 & (13) \\ 
1SWASP J052926.88+461147.5 & 0.2266 & 0.41 & 0.80 & 0.33 & 0.77 & 0.52 & 5077 & 5071 & (12) \\ 
NSVS 2175434               & 0.2210 & 0.33 & 0.81 & 0.27 & 0.80 & 0.51 & 4898 & 4903 & (14) \\ 
1SWASP J074658.62+224448.5 & 0.2208 & 0.35 & 0.79 & 0.28 & 0.80 & 0.52 & 4543 & 4717 & (14) \\ 
CC Com                     & 0.2207 & 0.53 & 0.71 & 0.37 & 0.70 & 0.53 & 4200 & 4300 & (15) \\ 
1SWASP J080150.03+471433.8 & 0.2175 & 0.43 & 0.72 & 0.32 & 0.71 & 0.49 & 4685 & 4696 & (16) \\ 
NSVS 7179685               & 0.2097 & 0.47 & 0.65 & 0.30 & 0.67 & 0.48 & 3979 & 4100 & (16) \\ 
SDSS J012119.10-001949.9   & 0.2052 & 0.50 & 0.51 & 0.26 & 0.61 & 0.45 & 3840 & 3812 & (17) \\ 
\hline
\end{tabular}}\label{tabCompare}
\begin{tabnote}
\footnotemark[$*$] 
 (1) \citet{2011MNRAS.415.1509L};  (2) \citet{2018AcA....68..449A};  (3) \citet{2017NewA...56...14T};  (4) \citet{2017RAA....17..115J};  (5) \citet{2020RAA....20...10Z};  (6) \citet{2019PASJ...71...39Z};  
 (7) \citet{2014A&A...563A..34L}:  (8) \citet{2016RAA....16..135K};  (9) \citet{djur16};
(10) \citet{2019PASJ...71...81S}; (11) \citet{2019RMxAA..55...65M}; (12) \citet{2018NewA...62...41K};  (13) \citet{2015AnA...578A.103L}; (14) \citet{2018PASA...35....8K}; (15) \citet{2011AN....332..626K}; (16) \citet{2015MNRAS.448.2890D}; (17) \citet{2015RAA....15.2237J}. \\
\end{tabnote}
\end{table}

\section{Conclusions}
\label{secConclusions}

We have analyzed, for the first time, the light curves of six short-period, totally eclipsing W UMa binaries observed by \citet{koen16}. The models we constructed fit the observations very well and yield orbital and stellar parameters similar to other short-period W UMa binaries in recent studies. Binaries at and under the short-period cutoff at 0.22 days are rare and this study contributes to our understanding of their peculiarities by increasing the size of the sample. 

Continued monitoring of our targets to gather eclipse timings would enable dedicated period studies and possible detections of a third body. It would also be beneficial to measure the radial velocities of these stars and compare their spectroscopic mass ratios with the photometric ones obtained in this study. This would help establish the reliability of automated q-search in totally eclipsing contact binaries, a topic of increasing importance as the amount of unutilized data from space telescopes and ground-based surveys continues to grow.

\begin{ack}
This research was funded by the Ministry of Education, Science and Technological Development of Republic of Serbia (contract No. 451-03-68/2020-14/200002). The authors thank the anonymous referee for useful suggestions that improved the paper. We also gratefully acknowledge the use of the Simbad database ({\it http://simbad.u-strasbg.fr/simbad/}), operated at the CDS, Strasbourg, France, and NASA's Astrophysics Data System Bibliographic Services ({\it http://adsabs.harvard.edu/})
\end{ack}

\end{document}